\documentclass[pra,twocolumn,floatfix,a4paper,superscriptaddress,showpacs]{revtex4}
\usepackage{amsfonts}
\usepackage{makeidx}
\usepackage{amssymb}
\usepackage{amsmath}
\usepackage{graphicx}
\usepackage{graphics}
\begin{document}
\title{Optimality of Gaussian discord}
\author{Stefano Pirandola}
\email{stefano.pirandola@york.ac.uk}
\affiliation{Computer Science, University of York, York YO10 5GH, United Kingdom}
\author{Gaetana Spedalieri}
\affiliation{Computer Science, University of York, York YO10 5GH, United Kingdom}
\author{Samuel L. Braunstein}
\affiliation{Computer Science, University of York, York YO10 5GH, United Kingdom}
\author{Nicolas J. Cerf}
\affiliation{Ecole Polytechnique de Bruxelles, CP 165, Universit\'{e} Libre de Bruxelles
(ULB), 1050 Brussels, Belgium}
\author{Seth Lloyd}
\affiliation{Research Laboratory of Electronics \& Department of Mechanical Engineering,
MIT, Cambridge, Massachusetts 02139, USA}

\begin{abstract}
In this Letter we exploit the recently-solved conjecture on the bosonic
minimum output entropy to show the optimality of Gaussian discord, so that the
computation of quantum discord for bipartite Gaussian states can be restricted
to local Gaussian measurements. We prove such optimality for a large family of
Gaussian states, including all two-mode squeezed thermal states, which are the
most typical Gaussian states realized in experiments. Our family also includes
other types of Gaussian states and spans their entire set in a suitable limit
where they become Choi-matrices of Gaussian channels. As a result, we
completely characterize the quantum correlations possessed by some of the most
important bosonic states in quantum optics and quantum information.

\end{abstract}

\maketitle

Quantum correlations represent a fundamental resource in quantum
information and computation~\cite{Nielsen,Wilde}. If we restrict
the description of a quantum system to pure states, then quantum
entanglement is synonymous with quantum correlations. However,
this is not exactly the case when general mixed states are
considered: Separable mixed states can still have residual
correlations which cannot be simulated by any classical
probability distribution~\cite{Qdiscord,Qdiscord2}. These residual
quantum correlations are today quantified by quantum
discord~\cite{VedralRMP}.

Quantum discord is defined as the difference between the total correlations
within a quantum state, as measured by the quantum mutual information, and its
classical correlations, corresponding to the maximal randomness which can be
shared by two parties by means of local measurements and one-way classical
communication~\cite{Winter}. This definition not only provides a more precise
characterization of quantum correlations but also has direct application in
various protocols, including quantum state merging~\cite{Merging1}, remote
state preparation~\cite{Remote}, discrimination of unitaries~\cite{Gu},
quantum channel discrimination~\cite{QILLdiscord}, quantum
metrology~\cite{Adesso} and quantum cryptography~\cite{QKDdiscord}.

For bosonic systems, like the optical modes of the electromagnetic field, it
is therefore crucial to compute the quantum discord of Gaussian
states~\cite{RMP}. Despite these states being the most common in experimental
quantum optics and the most studied in continuous-variable quantum
information~\cite{SamRev}, no closed formula is yet known for their quantum
discord. What is computed is an upper-bound, known as Gaussian
discord~\cite{GerryD,ParisD}, which is a simplified version based on Gaussian
detections only. Gaussian discord has been conjectured to be the actual
discord for Gaussian states, as also supported by recent numerical
studies~\cite{Paris,Olivares}.

In this Letter, we connect this conjecture on Gaussian discord with the
recently-solved conjecture on the bosonic minimum output entropy~\cite{Conje1}%
, according to which the von Neumann entropy at the output of a single-mode
Gaussian channel is minimized by a pure Gaussian state at the input~\cite{RMP}%
. In particular, this optimal input state is the vacuum, or any other coherent
state, when we consider Gaussian channels whose action is symmetric in the
quadratures (phase-insensitive), as for instance is the case of lossy or
amplifier channels (besides~\cite{Conje1} see also~\cite{Conje2} for an
alternate proof).

We show that the minimization of the bosonic output entropy implies the
optimality of Gaussian discord for a large family of Gaussian states. This
family includes the important class of squeezed thermal states, which are all
those states realized by applying two-mode squeezing to a pair of single-mode
thermal states~\cite{ParisD}. A key point in our analysis is showing that
these states can always be decomposed into an Einstein-Podolsky-Rosen (EPR)
state plus the local action of a phase-insensitive Gaussian channel. Given
such decomposition, we can easily show that heterodyne detection represents
the optimal local measurement for computing their quantum discord.

More generally, by extending the previous decomposition to include other forms
of local Gaussian channels, we show that we can generate many other types of
bipartite Gaussian states, for which the optimal local measurement is
Gaussian, given by a (quasi-)projection on single-mode pure squeezed states.
Furthermore, our decomposition spans the entire set of Gaussian states in a
suitable (and fastly-converging) limit where they become Choi-matrices of
local Gaussian channels.

As a result of our study, we are now able to compute the actual unrestricted
discord of a large portion of Gaussian states, paving the way for a complete
and precise characterization of the most fundamental quantum correlations
possessed by bosonic systems.

\textit{Quantum discord and its Gaussian formulation}.~~In classical
information theory, the mutual information between two random variables, $X$
and $Y$, can be written as $I(X,Y)=H(X)-H(X|Y)$, where $H(X)$ is the Shannon
entropy of variable $X$, and $H(X|Y)=H(X,Y)-H(Y)$ is its conditional Shannon
entropy. This notion has several inequivalent generalizations in quantum
information theory~\cite{VedralRMP}, where the two variables are replaced by
two quantum systems, $A$ and $B$, in a joint quantum state $\rho_{AB}$.

A first generalization is given by the quantum mutual information~\cite{Wilde}%
, defined as $I(A,B)=S(A)-S(A|B)$, where $S(A)=-\mathrm{Tr}(\rho_{A}\log
_{2}\rho_{A})$ is the von Neumann entropy of system $A$, in the reduced state
$\rho_{A}=\mathrm{Tr}_{B}(\rho_{AB})$, and $S(A|B)=S(A,B)-S(B)$ is the
conditional von Neumann entropy (to be computed from the joint state
$\rho_{AB}$ and the other reduced state $\rho_{B}$). The quantum mutual
information is a measure of the total correlations between the two quantum systems.

A second generalization is given by%
\begin{equation}
C(A|B)=S(A)-S_{\min}(A|B), \label{CDEVE}%
\end{equation}
where the conditional term $S_{\min}(A|B)$ is the von Neumann entropy of
system $A$ minimized over all possible measurements on system $B$, generally
described as positive operator valued measures (POVMs) $\mathcal{M}%
_{B}=\{M_{k}\}$ (in particular, it is sufficient to consider rank-1
POVMs~\cite{VedralRMP}). Mathematically, this conditional term is written as
\begin{equation}
S_{\min}(A|B):=\inf_{\mathcal{M}_{B}}S(A|\mathcal{M}_{B}), \label{minCOND}%
\end{equation}
where~\cite{Sum}%
\begin{equation}
S(A|\mathcal{M}_{B}):=\sum_{k}p_{k}S(\rho_{A|k}),
\end{equation}
with $p_{k}=\mathrm{Tr}(\rho_{AB}M_{k})$ the probability of outcome $k$, and
$\rho_{A|k}=p_{k}^{-1}\mathrm{Tr}_{B}(\rho_{AB}M_{k})$ the conditional state
of $A$.

The entropic quantity $C(A|B)$ quantifies the classical correlations in the
joint state $\rho_{AB}$, being the maximum amount of common randomness which
can be extracted by local measurements and one-way classical
communication~\cite{Winter}. Quantum discord is then defined as the difference
between the total and these classical
correlations~\cite{Qdiscord,Qdiscord2,VedralRMP}%
\begin{equation}
D(A|B):=I(A,B)-C(A|B)=S_{\min}(A|B)-S(A|B). \label{DISCORDeq}%
\end{equation}

When the two systems $A$ and $B$ are bosonic modes, we can consider their
Gaussian discord $D_{G}(A|B)\geq D(A|B)$, where the minimization of the
conditional term $S_{\min}(A|B)$ in Eq.~(\ref{minCOND}) is restricted to
Gaussian POVMs. Thanks to this restriction, Gaussian discord is easy to
compute for two-mode Gaussian states~\cite{GerryD,ParisD}. Furthermore, as we
have already mentioned, Gaussian discord is conjectured to be optimal for
these states, in the sense that it would represent their unrestricted quantum
discord, i.e., $D_{G}(A|B)=D(A|B)$. This conjecture is supported by numerical
studies~\cite{Paris,Olivares} and known to be true for a very limited set of
Gaussian states $\rho_{AB}$, namely those which can be purified into a
three-mode Gaussian state $\Phi_{ABE}$ symmetric in the $AE$
subsystem~\cite{GerryD,NOTEkw}.

\textit{Normal forms, two-mode squeezed thermal states, and their
decomposition}.~~Since quantum discord and classical correlations are entropic
quantities, they are invariant under local unitaries. This means that we may
apply displacements to yield a zero mean value, and local Gaussian unitaries
to reduce the covariance matrix (CM) into normal form~\cite{RMP}. Thus,
without loss of generality, quantum discord can be studied for zero-mean
Gaussian states $\rho_{AB}$ with CM%
\begin{equation}
\mathbf{V}_{AB}=\left(
\begin{array}
[c]{cc}%
a\mathbf{I} & \mathrm{diag}(c,c^{\prime})\\
\mathrm{diag}(c,c^{\prime}) & b\mathbf{I}%
\end{array}
\right)  :=\mathbf{V}(a,b,c,c^{\prime}), \label{normalFORM}%
\end{equation}
where $\mathbf{I}=\mathrm{diag}(1,1)$ and the parameters satisfy bona-fide
conditions imposed by the uncertainty principle~\cite{twoMODEs,pirNJP,Units}.

For simplicity, we start from the most typical zero-mean Gaussian states,
i.e., two-mode squeezed thermal states. These states have CMs of the form
$\mathbf{V}_{AB}=\mathbf{V}(a,b,c,-c)$ with bona-fide conditions $a,b\geq1$
and $c^{2}\leq ab-1-|a-b|$. As proven in the Supplemental Material and
depicted in Fig.~\ref{fig}, these states can always be decomposed as
$\rho_{AB}=(\mathcal{E}\otimes\mathcal{I})(\sigma_{aB})$, where $\mathcal{E}$
is a phase-insensitive Gaussian channel (details below), $\mathcal{I}$ is the
identity channel, and $\sigma_{aB}$ is an EPR state with CM%
\begin{equation}
\mathbf{V}_{aB}:=\left(
\begin{array}
[c]{cc}%
b\mathbf{I} & \sqrt{b^{2}-1}\mathbf{C}\\
\sqrt{b^{2}-1}\mathbf{C} & b\mathbf{I}%
\end{array}
\right)  , \label{CMgene}%
\end{equation}
where%
\begin{equation}
\mathbf{C}:=\left(
\begin{array}
[c]{cc}%
\mathrm{sign}(c) & 0\\
0 & -\mathrm{sign}(c)
\end{array}
\right)  .
\end{equation}
\begin{figure}[ptbh]
\vspace{-3.0cm}
\par
\begin{center}
\includegraphics[width=0.75\textwidth] {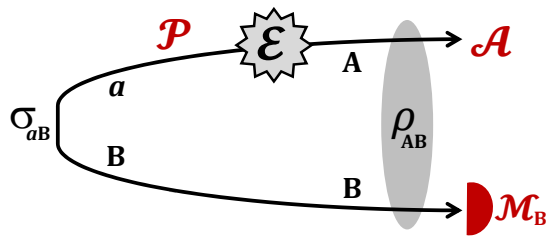}
\end{center}
\par
\vspace{-3.4cm}\caption{\textit{State decomposition}. A two-mode squeezed
thermal state $\rho_{AB}$ can always be decomposed into an EPR state
$\sigma_{aB}$ plus the local application of a phase-insensitive Gaussian
channel $\mathcal{E}$. \textit{Remote preparation} (red elements, see next
section). A local measurement $\mathcal{M}_{B}$ on mode $B$ generates a remote
ensemble $\mathcal{A}$\ on the output mode $A$. There will be another ensemble
$\mathcal{P}$ generated on the input mode $a$ before the channel. This input
ensemble will be made by Gaussianly-modulated coherent states if
$\mathcal{M}_{B}$ is heterodyne detection.}%
\label{fig}%
\end{figure}

At the level of the second order moments, a phase-insensitive Gaussian channel
$\mathcal{E}$\ performs the transformation $\mathbf{V}_{aB}\rightarrow
\mathbf{V}_{AB}=(\mathbf{K}\oplus\mathbf{I})\mathbf{V}_{aB}(\mathbf{K}%
^{T}\oplus\mathbf{I})+(\mathbf{N}\oplus\mathbf{0})$, with transmission matrix
$\mathbf{K}=\sqrt{\tau}\mathbf{I}$ and noise matrix $\mathbf{N}=\eta
\mathbf{I}$, where $\tau\geq0$ and $\eta\geq\left\vert 1-\tau\right\vert $. In
particular, this is a `lossy channel' for transmissivity $\tau\in\lbrack0,1]$
and thermal noise $\eta\geq1-\tau$, an `additive-noise channel' for $\tau=1$
and $\eta\geq0$, or an `amplifier channel' for $\tau>1$ and $\eta\geq\tau-1$.
(These are the most important canonical forms of a single-mode Gaussian
channel, see \cite{RMP} and Supplemental Material for this classification.) As
a result, the CM\ of a two-mode squeezed thermal state $\mathbf{V}%
_{AB}=\mathbf{V}(a,b,c,-c)$ can be expressed in the equivalent form%
\begin{equation}
\mathbf{V}_{AB}=\left(
\begin{array}
[c]{cc}%
(\tau b+\eta)\mathbf{I} & \sqrt{\tau(b^{2}-1)}\mathbf{C}\\
\sqrt{\tau(b^{2}-1)}\mathbf{C} & b\mathbf{I}%
\end{array}
\right)  , \label{CMab}%
\end{equation}
for some choice of $\tau\geq0$ and $\eta\geq|1-\tau|$.

\textit{Remote state preparation and connection with the bosonic minimum
output entropy}.~~As depicted in Fig.~\ref{fig}, the action of a local
POVM\ $\mathcal{M}_{B}=\{M_{k}\}$ on mode $B$ generates a remote ensemble of
states $\mathcal{P}$ on the input mode $a$, and a corresponding ensemble
$\mathcal{A}$ on the output mode $A$. With probability $p_{k}=\mathrm{Tr}%
(\sigma_{aB}M_{k})$, we have a conditional input state $\sigma_{a|k}%
=p_{k}^{-1}\mathrm{Tr}_{B}(\sigma_{aB}M_{k})\in\mathcal{P}$ and its
corresponding channel output $\rho_{A|k}=\mathcal{E}(\sigma_{a|k}%
)\in\mathcal{A}$.

Assuming heterodyne detection $\mathcal{M}_{B}=\mathrm{het}_{B}$, the input
ensemble $\mathcal{P}$ consists of coherent states $\sigma_{a|k}=|\alpha
_{k}\rangle\langle\alpha_{k}|$ whose complex amplitudes $\alpha_{k}$ are
Gaussianly-distributed (see Supplemental Material for more details on the
remote preparation of Gaussian states). As a result, the output ensemble
$\mathcal{A}$ will be composed of Gaussian states $\rho_{A|k}=\mathcal{E}%
(|\alpha_{k}\rangle\langle\alpha_{k}|)$ with Gaussianly-modulated first
moments and CM equal to $(\tau+\eta)\mathbf{I}$.

The output entropy associated with the heterodyne detection is equal to the
average entropy of the output ensemble $\mathcal{A}$, i.e.,%
\begin{equation}
S(A|\mathrm{het}_{B})=\int d^{2}k~p_{k}~S[\mathcal{E}(|\alpha_{k}%
\rangle\langle\alpha_{k}|)].
\end{equation}
Since entropy is invariant under displacements, we may write~\cite{Invariance}
$S[\mathcal{E}(|\alpha_{k}\rangle\langle\alpha_{k}|)]=S[\mathcal{E}%
(|0\rangle\langle0|)]$ and therefore%
\begin{equation}
S(A|\mathrm{het}_{B})=S[\mathcal{E}(|0\rangle\langle0|)]=h(\tau+\eta),
\end{equation}
where
\begin{equation}
h(x):=\frac{x+1}{2}\log_{2}\frac{x+1}{2}-\frac{x-1}{2}\log_{2}\frac{x-1}{2}.
\label{gVON}%
\end{equation}

At this point we can exploit the solved conjecture on the bosonic minimum
output entropy, which states that the vacuum (or any other coherent state)
minimizes the output entropy of the phase-insensitive Gaussian channel
$\mathcal{E}$ among all possible input states~\cite{Conje1,Conje2}
\begin{equation}
S[\mathcal{E}(|0\rangle\langle0|)]=\inf_{\rho}S[\mathcal{E}(\rho)].
\label{minOUT}%
\end{equation}
As a result, we may write%
\begin{align}
S(A|\mathrm{het}_{B})  &  =\inf_{\rho}S[\mathcal{E}(\rho)]\nonumber\\
&  \leq\inf_{\mathcal{M}_{B}}S(A|\mathcal{M}_{B})=S_{\min}(A|B),
\end{align}
where the inequality comes from the fact that any $\mathcal{M}_{B}$, with
input ensemble $\mathcal{P}=\{p_{k},\sigma_{a|k}\}$ and output ensemble
$\mathcal{A}=\{p_{k},\rho_{A|k}\}$, must satisfy
\begin{align}
S(A|\mathcal{M}_{B})  &  =\sum_{k}p_{k}S(\rho_{A|k})\geq\inf_{\mathcal{A}%
}S(\rho_{A|k})\nonumber\\
&  =\inf_{\mathcal{P}}S[\mathcal{E}(\sigma_{a|k})]\geq\inf_{\rho}%
S[\mathcal{E}(\rho)].
\end{align}

Thus, there is a Gaussian POVM (heterodyne detection) which is optimal for the
minimization of the output entropy $S(A|\mathcal{M}_{B})$. This is equivalent
to saying that the Gaussian discord of the Gaussian state $\rho_{AB}$\ is
optimal, i.e., equal to its actual discord. Its calculation is therefore easy,
since $S_{\min}(A|B)=h(\tau+\eta)$, leading to
\begin{equation}
D(A|B)=h(b)-h(\nu_{-})-h(\nu_{+})+h(\tau+\eta)~, \label{dFORMULA}%
\end{equation}
where $\{\nu_{\pm}\}$ is the symplectic spectrum of $\mathbf{V}_{AB}$, which
can be easily computed~\cite{RMP} from Eq.~(\ref{CMab}).

\textit{Extending the family of Gaussian states}.~Here we extend the previous
derivation to include other Gaussian states. We first generalize the local
Gaussian POVM $\mathcal{M}_{B}$, whose element $M_{k}$ becomes a
quasi-projector on the squeezed state $\left\vert k,u\right\rangle
$~\cite{RMP} with variable amplitude $k$ but fixed CM $\mathbf{V}_{\text{sq}%
}(u)=\mathrm{diag}(u,u^{-1})$, where $u>0$. This measurement $\mathcal{M}%
_{B}(u)$\ corresponds to a heterodyne detection for $u=1$, and becomes a
homodyne detection for $u\rightarrow0$ or $u\rightarrow+\infty$. By applying
$\mathcal{M}_{B}(u)$ to an EPR state $\sigma_{aB}$ with variance $b$ as
in\ Eq.~(\ref{CMgene}), we generate an ensemble $\mathcal{P}$ of
amplitude-modulated squeezed states with CM $\mathbf{V}_{\text{sq}}(r)$, where
$r=(1+ub)(u+b)^{-1}$. The value of this squeezing ranges from $r=b^{-1}$ to
$r=b$, extremes which are achieved by the two homodyne detectors (see
Supplemental Material for more details).

\begin{figure}[ptbh]
\vspace{-2.5cm}
\par
\begin{center}
\includegraphics[width=0.56\textwidth] {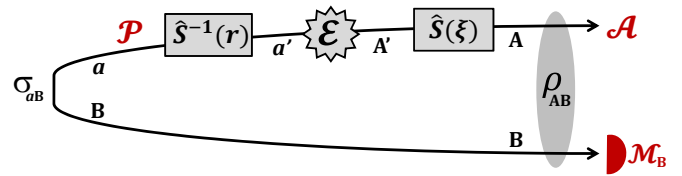}
\end{center}
\par
\vspace{-2.8cm}\caption{\textit{General decomposition}.~We consider Gaussian
states $\rho_{AB}$ which can be decomposed into an EPR state $\sigma_{aB}$ by
locally applying a Gaussian channel $\mathcal{E}$ plus input-output squeezing
operators. Here the local detection $\mathcal{M}_{B}$ is a quasi-projection
onto squeezed states, so that the input ensemble $\mathcal{P}$ is composed of
amplitude-modulated squeezed states.}%
\label{decoPIC}%
\end{figure}

We may now \textquotedblleft rectify\textquotedblright\ the ensemble
$\mathcal{P}$ by applying the anti-squeezing operator~\cite{RMP} $\hat{S}%
^{-1}(r)$ which transforms its states into coherent states (see
Fig.~\ref{decoPIC}). In this way we are sure that optimal states are fed into
the Gaussian channel $\mathcal{E}$, whose output entropy is therefore
minimized. Furthermore, this output entropy does not change if we apply
another squeezing operator $\hat{S}(\xi)$, with a suitable $\xi$ putting the
Gaussian state $\rho_{AB}$ into normal form. Thus the optimality of Gaussian
discord is proven for any Gaussian state which is decomposable into an
EPR\ state as
\begin{equation}
\rho_{AB}=(\mathcal{S}_{\xi}\mathcal{ES}_{r}^{-1}\otimes\mathcal{I}%
)(\sigma_{aB}),~\mathcal{S}_{x}(\rho):=\hat{S}(x)\rho\hat{S}^{\dagger}(x),
\label{genDEC}%
\end{equation}
with optimal detection given by the Gaussian\ $\mathcal{M}_{B}(u)$.

The second generalization consists of extending the Gaussian channel
$\mathcal{E}$\ to include negative transmissivities $\tau\leq0$, where the
channel describes the conjugate of an amplifier, which is another Gaussian
channel (phase-sensitive) whose output entropy is minimized by coherent states
at the input~\cite{Conje1,Conje2}. We can then consider an extended Gaussian
channel $\mathcal{E}$ with arbitrary $\tau\in\mathbb{R}$ and $\eta
\geq\left\vert 1-\tau\right\vert $, which is described by the matrices
$\mathbf{K}=\sqrt{|\tau|}\mathrm{diag}[1,\mathrm{sign}(\tau)]$ and
$\mathbf{N}=\eta\mathbf{I}$. Thus, the optimality of Gaussian discord is
proven for all Gaussian states $\rho_{AB}$ decomposable as in
Eq.~(\ref{genDEC}) with $\mathcal{E}$ being such an extended channel. As we
show in the Supplemental Material, these states have CMs in the normal-form
$\mathbf{V}(a,b,c,c^{\prime})$, where%
\begin{align}
a  &  =\theta(r)\theta(r^{-1}),~\theta(r):=\sqrt{\eta r+\left\vert
\tau\right\vert b},\label{a}\\
c  &  =\pm\sqrt{\left\vert \tau\right\vert (b^{2}-1)\theta(r^{-1})/\theta
(r)},\label{c}\\
c^{\prime}  &  =\mp\text{\textrm{sign}}[\tau]\sqrt{\left\vert \tau\right\vert
(b^{2}-1)\theta(r)/\theta(r^{-1})}, \label{cp}%
\end{align}
with $\tau\in\mathbb{R}$, $\eta\geq\left\vert 1-\tau\right\vert $ and
$r\in\lbrack b^{-1},b]$. Here the sign ambiguity comes from the type of EPR
state considered in the decomposition, whose CM in Eq.~(\ref{CMgene}) is
generally defined with $\mathbf{C}=\pm\mathrm{diag}(1,-1)$. Also note that
negative transmissivities allow us to include states with $cc^{\prime}\geq0$.

\begin{figure}[ptbh]
\vspace{+0.2cm}
\par
\begin{center}
\includegraphics[width=0.49\textwidth] {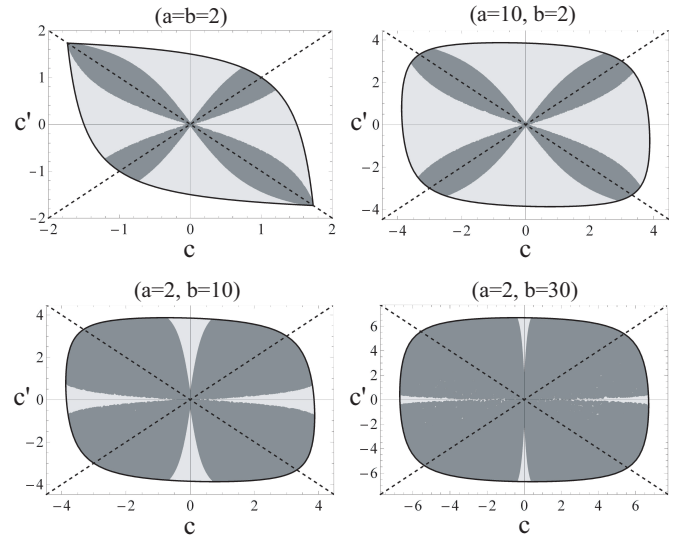}
\end{center}
\par
\vspace{-0.5cm}\caption{For given values of $a$ and $b$, the correlation
parameters $c$ and $c^{\prime}$ can only take a restricted range of physical
values corresponding to the delimited regions in the panels. Within these
regions the darker points are members of our Gaussian family. Plots are
created by randomly testing $5\times10^{5}$ points.}%
\label{picTOT}%
\end{figure}

For arbitrary fixed values of $a\geq1$ and $b\geq1$, we generate all
accessible values of the correlation parameters $c$ and $c^{\prime}$ by
exploiting the remaining degrees of freedom in the parameters $\tau$, $\eta$
and $r$. As shown in the numerical investigation of Fig.~\ref{picTOT}, our
family represents a wide portion of all Gaussian states. It includes all
states whose CMs have the form $\mathbf{V}(a,b,c,c^{\prime})$ with
$|c|=|c^{\prime}|$, corresponding to the bisectors of the correlation plane
$(c,c^{\prime})$. Furthermore, for increasing $b$, our family tends to invade
the entire Gaussian set very quickly. This set is completely filled for
$b\rightarrow+\infty$, where the EPR state $\sigma_{aB}$\ becomes maximally
entangled and the Gaussian state $\rho_{AB}$ becomes the Choi matrix of the
Gaussian channel $\mathcal{S}_{\xi}\mathcal{ES}_{r}^{-1}$.

By increasing the entanglement in the EPR state, we increase the amount of
squeezing $r$ that we can generate in the input ensemble $\mathcal{P}$. The
effect of this squeezing is to include output states where $|c|$ and
$|c^{\prime}|$ are very different, as also evident from the $r$-dependence in
Eqs.~(\ref{c}) and~(\ref{cp}). In particular, in the limit of $b\rightarrow
+\infty$, homodyne detectors generate infinite squeezing in $\mathcal{P}$ and
we approach all the exotic states on the axes of the plane, for which one of
the correlation parameters is zero ($cc^{\prime}=0$).

\textit{Conclusion and discussion}.~~In this Letter, we have shown that the
validity of the bosonic minimum output entropy conjecture implies the
optimality of Gaussian discord for a large family of Gaussian states,
decomposable into EPR states subject to local Gaussian channels. In
particular, this family includes the class of two-mode squeezed thermal
states~\cite{ParisD}. From this point of view, our work completely
characterizes the quantum correlations possessed by the most typical states in
continuous variable quantum information and experimental quantum optics. The
exact size of these quantum correlations, i.e., their unconstrained quantum
discord, can now be computed efficiently (note that computing discord is
NP-complete in the general case~\cite{Huang}).

We have shown that we can rapidly fill the entire set of Gaussian states by
increasing their parameter $b$, i.e., the thermal variance in the mode under
detection. The reason is because this parameter corresponds to the variance of
the EPR state\ involved in the decomposition. By increasing this parameter, we
increase the EPR entanglement and therefore the amount of squeezing that we
can remotely generate at the input of the Gaussian channel.

We could further improve our results if an energy-constrained version of the
entropy conjecture were proven for the `pathological' canonical
forms~\cite{RMP,HolevoCLASS,CanDILA}. In fact, there exist highly
phase-sensitive Gaussian channels which completely destroy the correlations in
only one of the quadratures, therefore being particularly suitable to
decompose exotic Gaussian states with $cc^{\prime}=0$. Unfortunately, no
finite-energy state is known to be optimal for the minimization of the output
entropy of these channels. If such a state were Gaussian, then it could be
prepared with a limited amount of squeezing and we would span the entire set
of Gaussian states more straightforwardly.

Finally, we remark that the complete characterization of the quantum
correlations possessed by Gaussian states is important not only in quantum
information and quantum optics (e.g., for problems of quantum metrology), but
also in other fields. These include condensed matter physics (e.g.,
Bose-Einstein condensates), solid-state physics, relativistic quantum field
theory (where Gaussian states arise from Bogoliubov transformations, e.g., in
the Unruh effect or the Hawking radiation), statistical mechanics and
foundations of quantum mechanics.

\textit{Acknowledgments}.~~This work was funded by a Leverhulme Trust research
fellowship and EPSRC (grants EP/J00796X/1 and EP/L011298/1). S.P. would like
to thank M.G.A. Paris for enlightening discussions.


\setcounter{section}{0} \setcounter{subsection}{0}
\renewcommand{\bibnumfmt}[1]{[S#1]} \renewcommand{\citenumfont}[1]{S#1}


\begin{center}
{\huge Supplemental Material}

\bigskip
\end{center}

\textbf{Contents of the document}.~In this Supplemental Material, we start by
giving a brief review on single-mode Gaussian channels and their canonical
forms, also discussing the minimization of their output entropy
(Sec.~\ref{canFORMS_app}). In Sec.~\ref{deco1_app} we prove the decomposition
of two-mode squeezed thermal states. In Sec.~\ref{deco2_app} we prove the
parametrization of our general family of Gaussian states and we describe the
random sampling of this family in more detail. In Sec.~\ref{Remote_app}, we
provide the mathematical tools for computing the remote preparation of
Gaussian states by applying a local Gaussian measurement to an arbitrary
two-mode Gaussian state. Formulas will be then specified to the case of the
EPR\ state, subject to the detections considered in the main article. Because
some of these tools are only partially known in the literature, we provide a
comprehensive proof in the final Sec.~\ref{APPmiscellanea}.

\section{Canonical forms of single-mode Gaussian channels\label{canFORMS_app}}

Any single-mode Gaussian channel $\mathcal{E}$ can be reduced to a simpler
`canonical form' up to Gaussian unitaries, applied at the input and the output
of the channel. The canonical forms are classified as $A_{1}$, $A_{2}$,
$B_{1}$, $B_{2}$, $C$ and $D$ (see Ref.~\cite{RMPapp,HolevoCLASSapp} for this
formalism). They transform the CM $\mathbf{V}$ of an input state as
$\mathbf{V}\rightarrow\mathbf{KVK}^{T}+\mathbf{N}$, with channel matrices
$\mathbf{K}$ and $\mathbf{N}$\ which are diagonal.

The most typical and important canonical forms are phase-insensitive, with
$\mathbf{K}$ and $\mathbf{N}$ both proportional to the identity, of the form
\begin{equation}
\mathbf{K}=\sqrt{\tau}\mathbf{I},~\mathbf{N}=\eta\mathbf{I},
\end{equation}
where $\tau$ is the transmissivity and $\eta$ is thermal noise. The first
example is the lossy channel (form $C$) for which $\tau\in\lbrack0,1]$ and
$\eta=(1-\tau)\omega$ with $\omega\geq1$. In terms of the quadrature operators
$\mathbf{\hat{x}}=(\hat{q},\hat{p})^{T}$ its action is expressed by the
input-output transformations%
\begin{equation}
\mathbf{\hat{x}}\rightarrow\sqrt{\tau}\mathbf{\hat{x}}+\sqrt{1-\tau
}\mathbf{\hat{x}}_{th},
\end{equation}
where $\mathbf{\hat{x}}_{th}$ are the quadratures of an environmental mode in
a thermal state with variance $\omega=2\bar{n}+1$, with $\bar{n}$ being the
mean number of thermal photons. At the border, for $\tau=0$, this channel
coincides with the form $A_{1}$ which is a completely depolarizing channel.
For $\tau=1$, it just becomes an identity channel (a particular instance of
the form $B_{2}$).

The second example is the additive-noise channel (form $B_{2}$), which has
$\tau=1$ and $\eta\geq0$. This channel corresponds to the transformation
\begin{equation}
\mathbf{\hat{x}}\rightarrow\mathbf{\hat{x}}+\boldsymbol{\xi},
\end{equation}
where $\boldsymbol{\xi}$ is a classical random variable with CM\ $\eta
\mathbf{I}$. The final example of phase-insensitive canonical form is the
amplifier channel (another kind of form $C$) which has $\tau\geq1$ and
$\eta=(\tau-1)\omega$. This is realized by
\begin{equation}
\mathbf{\hat{x}}\rightarrow\sqrt{\tau}\mathbf{\hat{x}}+\sqrt{\tau
-1}\mathbf{\hat{x}}_{th},
\end{equation}
where $\mathbf{\hat{x}}_{th}$ are the quadratures of a thermal mode.

Then, we have canonical forms which are sensitive to phase.\ The most
important is the form $D$ which is the conjugate of the amplifier channel,
i.e., the environmental output of a two-mode squeezer which dilates the
amplifier channel. This form can be associated with negative transmissivities
$\tau\leq0$ and is described by the matrices
\[
\mathbf{K}=\sqrt{-\tau}\mathbf{Z},~\mathbf{N}=\eta\mathbf{I},
\]
with $\mathbf{Z}:=\mathrm{diag}(1,-1)$ and $\eta=(1-\tau)\omega=(1-\tau
)(2\bar{n}+1)$. Its sensitivity to phase is clear from the fact that
$\mathbf{K}\propto\mathrm{diag}(1,-1)$. It corresponds to the transformation%
\begin{equation}
\mathbf{\hat{x}}\rightarrow\sqrt{-\tau}\mathbf{Z\hat{x}}+\sqrt{1-\tau
}\mathbf{\hat{x}}_{th}.
\end{equation}

In the main paper, our phase-insensitive Gaussian channel is given by the
forms $C$ and $B_{2}$, while our extended Gaussian channel is given by the
forms $C$, $B_{2}$ and $D$. Such an extended channel can be compactly
characterized by the matrices%
\begin{equation}
\mathbf{K}=\sqrt{|\tau|}\mathrm{diag}[1,\mathrm{sign}(\tau)],~\mathbf{N}%
=\eta\mathbf{I},
\end{equation}
with $\tau\in\mathbb{R}$ and $\eta\geq\left\vert 1-\tau\right\vert $. The
corresponding input-output relations can be written in the form%
\begin{equation}
\mathbf{\hat{x}}\rightarrow\left(
\begin{array}
[c]{cc}%
\sqrt{|\tau|} & 0\\
0 & [\mathrm{sign}(\tau)]\sqrt{|\tau|}%
\end{array}
\right)  \mathbf{\hat{x}}+\mathbb{\zeta}, \label{classN}%
\end{equation}
where $\mathbb{\zeta}$ is a noise-variable with CM equal to $\eta\mathbf{I}$.
The output entropy of this extended Gaussian channel is minimized by taking a
coherent state at the input~\cite{Conje1app}.

Finally, we have the `pathological' canonical forms $A_{2}$ and $B_{1}$, which
are highly phase-sensitive. In particular, form $A_{2}$ is described by
\begin{equation}
\mathbf{K}=\mathrm{diag}(1,0),~\mathbf{N}=(2\bar{n}+1)\mathbf{I},
\end{equation}
while form $B_{1}$ is described by
\begin{equation}
\mathbf{K}=\mathbf{I},~\mathbf{N}=\mathrm{diag}(0,1).
\end{equation}
Form $A_{2}$ is highly symmetric in the transmission matrix, and represents a
sort of `half' depolarizing channel which replaces one of the input
quadratures with thermal noise. When this form is locally applied to a
bipartite state, it completely destroys the correlations in one of the
quadratures. By contrast, $B_{1}$ is highly asymmetric in the noise matrix and
its effect is to add a unit of vacuum noise to only one of the quadratures.

For these pathological forms, the optimal input state is expected to be a
squeezed vacuum. Indeed, this can easily be proven for $B_{1}$, for which the
optimal input state is an infinitely-squeezed vacuum, with CM $\mathbf{V}%
=\mathrm{diag}(r^{-1},r)$ where $r\rightarrow+\infty$. In fact, at the output
we have a Gaussian state with covariance matrix\ $\mathbf{V}+\mathbf{N}%
=\mathrm{diag}(r^{-1},r+1)$ whose determinant goes to $1$, so that it is
asympotically pure. This implies that the output entropy goes to zero, i.e.,
to its minimum value. However, if we enforce an energy constraint at the input
of this form, then we do not know the optimal input state which minimizes the
output entropy.

The situation is more involved for the other form $A_{2}$. When $\bar{n}=0$,
this is the conjugate channel of $B_{1}$ in a unitary dilation with a pure
environment (Stinespring dilation). Since the global output of this dilation
is pure for an input pure state (we can always restrict the search to pure
states due to the concavity of the entropy), we have that the output entropy
of the conjugate channel $A_{2}$ is equal to the output entropy of $B_{1}$.
This trivially implies that the infinitely-squeezed vacuum is again the
optimal input state. However, for $\bar{n}>0$, the dilation involves a thermal
environment, and the optimality of this state is only conjectured for $A_{2}$.
Furthermore, also for this form, the optimal input is not known if we enforce
an energy constraint.

\section{Decomposition of two-mode squeezed thermal states\label{deco1_app}}

Here we prove that a CM\ in the special normal-form $\mathbf{V}(a,b,c,-c)$ can
equivalently be expressed as in Eq.~(\ref{CMab}), i.e., using the
parametrization
\begin{equation}
a=\tau b+\eta,~c=\sqrt{\tau(b^{2}-1)}, \label{paraAPP}%
\end{equation}
for some choice of $\tau\geq0$ and $\eta\geq|1-\tau|$.

Let us start by re-writing the bona-fide conditions for the parameters $a$,
$b$, and $c$. From the uncertainty principle, we derive~\cite{twoMODEsapp}
$a,b\geq1$ and $c^{2}\leq c_{\max}^{2}$, where%
\begin{align}
c_{\max}^{2}  &  :=\min\{(a-1)(b+1),(a+1)(b-1)\}\\
&  =ab-1-|a-b|.
\end{align}
For $a=1$ or $b=1$ we must have $c=0$. This CM\ is just realized by taking
$\tau=0$. For any $a,b>1$, it is sufficient to prove that we can generate all
CMs with maximal correlations $c^{2}=c_{\max}^{2}$ by means of our
parametrization. If this is possible for maximal correlations, then it will
also be possible for intermediate correlations $c^{2}\leq c_{\max}^{2}$, where
conditions are less stringent (since this extension is merely technical, it
has been omitted).

First consider the case $a\geq b>1$, so that $c_{\max}^{2}=(a+1)(b-1)$.
Imposing $c^{2}=(a+1)(b-1)$ and using Eq.~(\ref{paraAPP}), we derive
$\eta=\tau-1$ (possible for $\tau\geq1$). Using $\tau=\eta+1$ in the
parametrization of $a$, we find $a=b+\eta(1+b)$, which means that we can
generate all possible values of $a\geq b$ by freely varying $\eta\geq0$.

Consider now $b\geq a>1$, so that $c_{\max}^{2}=(a-1)(b+1)$. By imposing
$c^{2}=(a-1)(b+1)$ and using Eq.~(\ref{paraAPP}), we derive $\eta=1-\tau$,
which implies $0\leq\eta\leq1$. By replacing $\tau=1-\eta$ in the
parametrization of $a$, we derive $a=b(1-\eta)+\eta$. It is clear that we can
generate all possible values of $a\leq b$ by freely varying $0\leq\eta<1$.

\section{Parametrization of our family of Gaussian states\label{deco2_app}}

In order to derive the parametrization given in Eqs.~(\ref{a})-(\ref{cp}), we
work in the Heisenberg picture, considering the input-output transformations
for the quadrature operators $\mathbf{\hat{x}}=(\hat{q},\hat{p})^{T}$. The
action of the first squeezer in Fig.~\ref{decoPIC} is given by%
\begin{equation}
\mathbf{\hat{x}}_{a}\rightarrow\mathbf{\hat{x}}_{a^{\prime}}=S^{-1}%
(r)\mathbf{\hat{x}}_{a},
\end{equation}
where $r$ is positive and
\begin{equation}
S(r):=\left(
\begin{array}
[c]{cc}%
r^{1/2} & 0\\
0 & r^{-1/2}%
\end{array}
\right)  .
\end{equation}
The role of this squeezer is to rectify the input ensemble created by the
local Gaussian POVM $\mathcal{M}_{B}(u)$ with $u>0$, in such a way to have
coherent states on mode $a^{\prime}$. In fact, applying this local detection
to the $B$ mode of an EPR state $\sigma_{aB}$ with CM%
\begin{equation}
\mathbf{V}_{aB}=\left(
\begin{array}
[c]{cc}%
b\mathbf{I} & \sqrt{b^{2}-1}\mathbf{C}\\
\sqrt{b^{2}-1}\mathbf{C} & b\mathbf{I}%
\end{array}
\right)  ,~\mathbf{C=\pm}\left(
\begin{array}
[c]{cc}%
1 & 0\\
0 & -1
\end{array}
\right)  ,
\end{equation}
the other mode $a$ is projected onto amplitude-modulated squeezed states with
squeezing $r=(1+ub)(u+b)^{-1}$ which ranges between $b^{-1}$ and $b$ (see
Sec.~\ref{Remote_app} for more details on the remote preparation of Gaussian states).

After the first squeezer, the action of the extended Gaussian channel is
expressed by
\begin{equation}
\mathbf{\hat{x}}_{a^{\prime}}\rightarrow\mathbf{\hat{x}}_{A^{\prime}}=\left(
\begin{array}
[c]{cc}%
\sqrt{|\tau|} & 0\\
0 & [\mathrm{sign}(\tau)]\sqrt{|\tau|}%
\end{array}
\right)  \mathbf{\hat{x}}_{a^{\prime}}+\mathbb{\zeta~},
\end{equation}
where $\mathbb{\zeta}$ is a noise-variable with CM$\ \eta\mathbf{I}$, where
$\eta\geq|1-\tau|$ and $\tau\in\mathbb{R}$ (see Eq.~(\ref{classN}) in
Sec.~\ref{canFORMS_app}). Finally, the second squeezer realizes%
\begin{equation}
\mathbf{\hat{x}}_{A^{\prime}}\rightarrow\mathbf{\hat{x}}_{A}=S(\xi
)\mathbf{\hat{x}}_{A^{\prime}},
\end{equation}
where parameter $\xi$ is chosen in such a way to put the output
CM\ $\mathbf{V}_{AB}$ in normal-form. After simple algebra we find
\begin{equation}
\xi=r\frac{\theta(r^{-1})}{\theta(r)},~~\theta(r):=\sqrt{\eta r+\left\vert
\tau\right\vert b}. \label{theta_app}%
\end{equation}

Using these equations and the fact that mode $B$ is subject to the identity
(before detection), we have that the input EPR\ state is mapped into the
output Gaussian state $\rho_{AB}$ with normal-form CM $\mathbf{V}%
_{AB}=\mathbf{V}(a,b,c,c^{\prime})$, where%
\begin{align}
a  &  =\theta(r)\theta(r^{-1})\label{a_app}\\
c  &  =\pm\sqrt{\left\vert \tau\right\vert (b^{2}-1)\frac{\theta(r^{-1}%
)}{\theta(r)}},\label{c_app}\\
c^{\prime}  &  =\mp\text{\textrm{sign}}[\tau]\sqrt{\left\vert \tau\right\vert
(b^{2}-1)\frac{\theta(r)}{\theta(r^{-1})}}, \label{cp_app}%
\end{align}
which correspond to Eqs.~(\ref{a})-(\ref{cp}).

In order to generate elements of the family, we randomly pick values for
$b\geq1$, $r\in\lbrack b^{-1},b]$, $\tau\in\mathbb{R}$ and $\eta\geq|1-\tau|$.
Then, we compute $a$, $c$, and $c^{\prime}$ according to Eqs.~(\ref{a_app}%
)-(\ref{cp_app}). Alternatively, we can pick values for $a,b\geq1$ and
randomly generate values for $r$, $\tau$, and $\eta$ which are compatible with
Eq.~(\ref{a_app}). Using these values, we finally compute $c$ and $c^{\prime}$
according to Eqs.~(\ref{c_app}) and~(\ref{cp_app}). We adopt this second
procedure since it allows us to represent the elements of the family on the
correlation plane $(c,c^{\prime})$ for given values of $a$ and $b$.

Fixing a value for $a\geq1$, the condition in Eq.~(\ref{a_app}) restricts the
range of variability of the transmissivity. In fact, we can express the
parameter $\eta$ as function of $a$ by inverting Eq.~(\ref{a_app}) and finding%
\begin{equation}
\eta=\frac{\sqrt{4a^{2}r^{2}+(r^{2}-1)^{2}\tau^{2}b^{2}}-(1+r^{2})|\tau|b}%
{2r}. \label{eta_app}%
\end{equation}
Then, imposing $\eta\geq|1-\tau|$ implies $\tau\in\lbrack\tau_{\min}%
,\tau_{\max}]$, where%
\begin{align}
\tau_{\min}  &  =\frac{b+(br+2)r-\sqrt{(r^{2}-1)^{2}b^{2}+4a^{2}r\gamma_{+}}%
}{2\gamma_{+}},\\
\tau_{\max}  &  =\left\{
\begin{array}
[c]{c}%
\frac{b+(br-2)r-\sqrt{(r^{2}-1)^{2}b^{2}-4a^{2}r\gamma_{-}}}{2\gamma_{-}%
}~~\text{for }a\leq b,\\
\\
\frac{b+(br+2)r+\sqrt{(r^{2}-1)^{2}b^{2}+4a^{2}r\gamma_{+}}}{2\gamma_{+}%
}~~\text{for }b\leq a,
\end{array}
\right.
\end{align}
and $\gamma_{\pm}:=(r\pm b)(rb\pm1)$. Thus, for any values of $a,b\geq1$, we
randomly generate the parameters $r\in\lbrack b^{-1},b]$ and $\tau\in
\lbrack\tau_{\min},\tau_{\max}]$. Then, we compute $\eta$ according to
Eq.~(\ref{eta_app}) to be replaced in $\theta$ of Eq.~(\ref{theta_app}). We
finally compute the values of $c$ and $c^{\prime}$ according to
Eqs.~(\ref{c_app}) and~(\ref{cp_app}).

\section{Remote preparation of Gaussian states\label{Remote_app}}

Let us consider an arbitrary two-mode Gaussian state $\rho_{AB}$ with mean
value $\mathbf{\bar{x}}_{AB}^{T}=(\mathbf{\bar{x}}_{A}^{T},\mathbf{\bar{x}%
}_{B}^{T})\in\mathbb{R}^{4}$ and CM
\begin{equation}
\mathbf{V}_{AB}=\left(
\begin{array}
[c]{cc}%
\mathbf{A} & \mathbf{C}\\
\mathbf{C}^{T} & \mathbf{B}%
\end{array}
\right)  , \label{VabBlockFORM}%
\end{equation}
where $\mathbf{A=A}^{T}$, $\mathbf{B=B}^{T}$ and $\mathbf{C}$ are $2\times2$
real blocks. Mode $B$ is detected by an arbitrary Gaussian measurement, which
is defined as a POVM with measurement operator
\begin{equation}
M_{k}:=\pi^{-1}\hat{D}(k)\rho_{0}\hat{D}(-k), \label{meas_operator}%
\end{equation}
where $k=(q+ip)/2$ is the complex outcome, $\hat{D}(k)=\exp(k^{\ast}\hat
{a}-k\hat{a}^{\dagger})$ is the displacement operator~\cite{Walls} with
annihilation operator $\hat{a}=(\hat{q}+i\hat{p})/2$, and $\rho_{0}$ is a
zero-mean Gaussian state with CM $\mathbf{V}_{0}$. This operator displaces the
state by $-k$ and projects it onto the Gaussian state $\rho_{0}$. The Gaussian
measurement is rank-one when $\rho_{0}$ is pure. The simplest choice is the
vacuum state $\rho_{0}=\left\vert 0\right\rangle \left\langle 0\right\vert $,
in which case the Gaussian measurement corresponds to heterodyne detection.

The Gaussian POVM can also be expressed in terms of the quadrature operators
$\mathbf{\hat{x}}^{T}:=(\hat{q},\hat{p})$ and the real vector $\mathbf{k}%
^{T}=(q,p)$. In fact, from the decomposition of the identity%
\[
\hat{I}=\int_{\mathbb{C}}d^{2}k~M_{k}=\int_{\mathbb{R}^{2}}d^{2}%
\mathbf{k}~M_{\mathbf{k}},
\]
we derive the equivalent measurement operator%
\[
M_{\mathbf{k}}:=\frac{1}{4\pi}\rho_{0}(\mathbf{k}),
\]
where $\rho_{0}(\mathbf{k}):=\hat{D}(\mathbf{k})\rho_{0}\hat{D}(-\mathbf{k})$
and%
\begin{equation}
\hat{D}(\mathbf{k})=\exp\left(  \frac{i}{2}~\mathbf{\hat{x}}^{T}%
\boldsymbol{\Omega~}\mathbf{k}\right)  ,~\boldsymbol{\Omega}:=\left(
\begin{array}
[c]{cc}%
0 & 1\\
-1 & 0
\end{array}
\right)  . \label{Weyl1}%
\end{equation}

By detecting mode $B$, the outcome $\mathbf{k}$ is achieved with probability
\begin{equation}
p(\mathbf{k})=\frac{\exp\left[  -\frac{1}{2}\mathbf{d}^{T}(\mathbf{B}%
+\mathbf{V}_{0})^{-1}\mathbf{d}\right]  }{2\pi\sqrt{\det(\mathbf{B}%
+\mathbf{V}_{0})}},~\mathbf{d}:=\mathbf{\bar{x}}_{B}-\mathbf{k},
\label{px_for}%
\end{equation}
which is Gaussian with classical CM $\mathbf{V}_{\mathbf{k}}=\mathbf{B}%
+\mathbf{V}_{0}$. Correspondingly, the other mode $A$ is projected into a
conditional Gaussian state $\rho_{A|\mathbf{k}}$ with mean value%
\begin{equation}
\mathbf{\bar{x}}_{A|\mathbf{k}}=\mathbf{\bar{x}}_{A}-\mathbf{C}(\mathbf{B}%
+\mathbf{V}_{0})^{-1}\mathbf{d}, \label{xcond_for}%
\end{equation}
and CM given by
\begin{equation}
\mathbf{V}_{A|\mathbf{k}}=\mathbf{A}-\mathbf{C}(\mathbf{B}+\mathbf{V}%
_{0})^{-1}\mathbf{C}^{T}. \label{CMcond_for}%
\end{equation}

If we detect mode $A$ with outcome $\mathbf{k}$, then we have to permute $A$
and $B$ in the formulas. This means that $\mathbf{\bar{x}}_{A|\mathbf{k}}$ and
$\mathbf{V}_{A|\mathbf{k}}$ are replaced by $\mathbf{\bar{x}}_{B|\mathbf{k}}$
and $\mathbf{V}_{B|\mathbf{k}}$. Then, Eqs.~(\ref{px_for}), (\ref{xcond_for})
and~(\ref{CMcond_for}), are subject to the replacements $\mathbf{\bar{x}}%
_{A}\rightarrow\mathbf{\bar{x}}_{B}$, $\mathbf{\bar{x}}_{B}\rightarrow
\mathbf{\bar{x}}_{A}$, $\mathbf{A}\rightarrow\mathbf{B},$ $\mathbf{B}%
\rightarrow\mathbf{A}$ and $\mathbf{C}\rightarrow\mathbf{C}^{T}$.

Despite the fact that Eq.~(\ref{CMcond_for}) is well-known in the
literature~\cite{JensReview,RMPapp,giedkeapp}, the other two
Eqs.~(\ref{px_for}) and~(\ref{xcond_for}) are rarely-used and almost unknown.
For the sake of completeness and pedagogical reasons we prove all three
formulas in Sec.~\ref{APPmiscellanea}.

\subsection{Coherent state preparation with EPR states}

For its relevance to our work, we consider the specific case where the
two-mode Gaussian state $\rho_{AB}$ is an EPR\ state, and mode $B$ is
heterodyned, so that coherent states are prepared on mode $A$. Let us start
with one type of EPR\ state, whose CM $\mathbf{V}_{AB}$ in
Eq.~(\ref{VabBlockFORM}) has blocks
\[
\mathbf{A}=\mathbf{B}=\mu\mathbf{I},~\mathbf{C}=\sqrt{\mu^{2}-1}\mathbf{Z},
\]
where $\mu\geq1$ and $\mathbf{Z}=\mathrm{diag}(1,-1)$. Afterwards we deal with
the other EPR\ type ($\mathbf{Z}\rightarrow-\mathbf{Z}$).

Since mode $B$ is heterodyned, we have $\mathbf{V}_{0}=\mathbf{I}$ in the
previous formulas. According to Eq.~(\ref{px_for}), the outcome $\mathbf{k}$
follows a Gaussian probability $p(\mathbf{k})$ with classical CM
$\mathbf{V}_{\mathbf{k}}=(\mu+1)\mathbf{I}$, as expected since the reduced
state $\rho_{B}$ is thermal with CM $\mu\mathbf{I}$. Correspondingly, the
other EPR mode $A$ is projected into a conditional Gaussian state
$\rho_{A|\mathbf{k}}$ with mean-value
\begin{equation}
\mathbf{\bar{x}}_{A|\mathbf{k}}:=\mathbf{a}=\frac{\sqrt{\mu^{2}-1}}{\mu
+1}\mathbf{Zk}, \label{cohprepa}%
\end{equation}
from Eq.~(\ref{xcond_for}), and CM $\mathbf{V}_{A|\mathbf{k}}=\mathbf{I}$,
from Eq. (\ref{CMcond_for}). It is therefore a coherent state with a
Gaussianly-modulated mean-value $\mathbf{a}=\mathbf{a}(\mathbf{k})$. In
particular, the Gaussian distribution $p(\mathbf{a})$ has CM $(\mu
-1)\mathbf{I}$, as evident from the decomposition of the average (thermal)
state%
\[
\rho_{A}=\int d^{2}\mathbf{k}~p(\mathbf{k})\left\vert \mathbf{a}%
(\mathbf{k})\right\rangle \left\langle \mathbf{a}(\mathbf{k})\right\vert =\int
d^{2}\mathbf{a}~p(\mathbf{a})\left\vert \mathbf{a}\right\rangle \left\langle
\mathbf{a}\right\vert ,
\]
where%
\[
p(\mathbf{a})=\frac{\mu+1}{\mu-1}p[\mathbf{k}(\mathbf{a})]=\frac{\exp\left(
\frac{-\mathbf{a}^{T}\mathbf{a}}{2(\mu-1)}\right)  }{2\pi(\mu-1)}.
\]
In the complex notation, the amplitude of the remote coherent state is given
by
\begin{equation}
\alpha=\frac{\sqrt{\mu^{2}-1}}{\mu+1}k^{\ast},
\end{equation}
where $k$ is the complex outcome of the heterodyne detection. For large $\mu$,
we have $\mathbf{a}\rightarrow\mathbf{Zk}$ and $\alpha\rightarrow k^{\ast}$.

Consider now the other type of EPR\ state, i.e., the one having $\mathbf{C}%
=-\sqrt{\mu^{2}-1}\mathbf{Z}$. From Eqs.~(\ref{px_for})-(\ref{CMcond_for}), we
see that the sign change in $\mathbf{C}$ only affects the first moments in
Eq.~(\ref{xcond_for}). This means that mode $A$ is projected on coherent
states as before, the only difference being the relation between their
amplitude and the outcome of the heterodyne detection, which is now%
\begin{equation}
\mathbf{\bar{x}}_{A|\mathbf{k}}=\frac{-\sqrt{\mu^{2}-1}}{\mu+1}\mathbf{Zk}.
\label{coheprepa2}%
\end{equation}

Thus, heterodyning mode $B$ of an arbitrary EPR\ state with $\mathbf{C}%
=\pm\sqrt{\mu^{2}-1}\mathbf{Z}$ projects mode $A$ into an ensemble of coherent
states with coherent amplitudes given by Eq.~(\ref{cohprepa}) or
Eq.~(\ref{coheprepa2}). In terms of entropy there is no difference, since the
entropy of a Gaussian state depends only on the second-order statistical moments.

\subsection{Squeezed-state preparation with EPR states}

Here we show how we can create an ensemble of squeezed states by detecting one
mode of an EPR\ state. To achieve this, the Gaussian state $\rho_{0}$ in the
measurement operator $M_{k}$ of Eq.~(\ref{meas_operator}) must be the squeezed
vacuum, with CM $\mathbf{V}_{\text{sq}}(u)=\mathrm{diag}(u,u^{-1})$ where
$u>0$. Let us apply this local detection to mode $B$ of an EPR\ state with
blocks $\mathbf{A}=\mathbf{B}=\mu\mathbf{I}$ and $\mathbf{C}=\pm\sqrt{\mu
^{2}-1}\mathbf{Z}$. Using Eq.~(\ref{CMcond_for}) we derive
\begin{equation}
\mathbf{V}_{A|\mathbf{k}}=\left(
\begin{array}
[c]{cc}%
r & 0\\
0 & r^{-1}%
\end{array}
\right)  ,~r:=\frac{1+u\mu}{u+\mu}.
\end{equation}

This is clearly the CM\ of a squeezed state. In particular, it is easy to
check that for $u\rightarrow0$ (corresponding to homodyne $q$-detection), we
get $r\rightarrow\mu^{-1}$, while for $u\rightarrow+\infty$ (which is homodyne
$p$-detection) we get $r\rightarrow\mu$. The amount of squeezing we can
generate on mode $A$ is limited by the amount of correlations in the EPR state
(quantified by the parameter $\mu$). Finally, note that for $u=1$, we have
heterodyne detection and we get $r=1$ as expected (coherent states).

Finally, using Eqs.~(\ref{px_for}) and~(\ref{xcond_for}) we can derive both
the Gaussian probability of the outcomes and the modulated mean value of the
states in terms of these outcomes. However, we omit here this calculation,
being irrelevant for the computation of the entropy.

\section{Gaussian preparation: Proofs\label{APPmiscellanea}}

For the sake of completeness, we prove here the formulas~(\ref{px_for}%
)-(\ref{CMcond_for}) for the remote preparation of Gaussian states. We start
by introducing the partial-trace rule, which is a central tool in our demonstration.

\subsection{Partial-trace rule}

Consider\ a two-mode state $\rho_{AB}$ with characteristic
function~\cite{Walls} $\chi(\alpha,\beta):=\mathrm{Tr}[\rho_{AB}\hat{D}%
(\alpha)\otimes\hat{D}(\beta)]$, and a state $\rho_{0}$ of mode $B$ with
characteristic function $\chi_{0}(\beta):=\mathrm{Tr}[\rho_{0}\hat{D}(\beta
)]$. We consider the partial trace of their product, i.e., the operator
$\hat{O}:=\mathrm{Tr}_{B}[\rho_{AB}(I\otimes\rho_{0})]$, with associated
characteristic function $\chi\lbrack\hat{O}](\alpha):=\mathrm{Tr}[\hat{O}%
\hat{D}(\alpha)]$. This function can be computed via the partial-trace
rule~\cite{Proof}%
\begin{equation}
\chi\lbrack\hat{O}](\alpha)=\int_{\mathbb{C}}\frac{d^{2}\beta}{\pi}\chi
(\alpha,-\beta)\chi_{0}(\beta). \label{chiOB}%
\end{equation}
This is a simple extension of the trace rule for single-mode states $\rho$ and
$\rho_{0}$, which reads%
\begin{equation}
\mathrm{Tr}\left(  \rho\rho_{0}\right)  =\int_{\mathbb{C}}\frac{d^{2}\beta
}{\pi}\chi(-\beta)\chi_{0}(\beta).
\end{equation}
In particular, when $\rho$ and $\rho_{0}$ are Gaussian with statistical
moments $\{\mathbf{\bar{x}},\mathbf{V}\}$ and $\{\mathbf{\bar{x}}%
_{0},\mathbf{V}_{0}\}$, it is straightforward to check that%
\begin{equation}
\mathrm{Tr}\left(  \rho\rho_{0}\right)  =\frac{2\exp\left[  -\frac{1}%
{2}(\mathbf{\bar{x}-\bar{x}}_{0})^{T}(\mathbf{V}+\mathbf{V}_{0})^{-1}%
(\mathbf{\bar{x}-\bar{x}}_{0})\right]  }{\sqrt{\det(\mathbf{V}+\mathbf{V}%
_{0})}}. \label{traceRULEgaussian}%
\end{equation}

\subsection{Proof of Eqs.~(\ref{px_for})-(\ref{CMcond_for})}

Let us start with the proof of Eq.~(\ref{px_for}). We write%
\begin{equation}
p(\mathbf{k})=\mathrm{Tr}(\rho_{B}M_{\mathbf{k}})=\frac{1}{4\pi}%
\mathrm{Tr}\left[  \rho_{B}\rho_{0}(\mathbf{k})\right]  , \label{eqPa}%
\end{equation}
where $\rho_{B}=\mathrm{Tr}_{A}(\rho_{AB})$ is the reduced state of mode $B$.
Since $\rho_{B}$ and $\rho_{0}(\mathbf{k})$ are single-mode Gaussian states
with statistical moments $\{\mathbf{\bar{x}}_{B},\mathbf{B}\}$ and
$\{\mathbf{k},\mathbf{V}_{0}\}$, we can directly apply
Eq.~(\ref{traceRULEgaussian}) to obtain Eq.~(\ref{px_for}).

To compute the statistical moments $\mathbf{\bar{x}}_{A|\mathbf{k}}$ and
$\mathbf{V}_{A|\mathbf{k}}$ of the conditional Gaussian state
\[
\rho_{A|\mathbf{k}}=p(\mathbf{k})^{-1}\mathrm{Tr}_{B}(\rho_{AB}M_{\mathbf{k}%
})=\frac{1}{4\pi p(\mathbf{k})}\mathrm{Tr}_{B}\left[  \rho_{AB}\rho
_{0}(\mathbf{k})\right]  ,
\]
we apply the partial-trace rule to $\hat{O}_{\mathbf{k}}:=\mathrm{Tr}%
_{B}\left[  \rho_{AB}\rho_{0}(\mathbf{k})\right]  $. Here it is convenient to
adopt the Cartesian decomposition $\hat{a}=(\hat{q}+i\hat{p})/2$ and
$\alpha=(q+ip)/2$, and use the following real variables
\[
\mathbf{\hat{x}}:=\left(
\begin{array}
[c]{c}%
\hat{q}\\
\hat{p}%
\end{array}
\right)  ,~\mathbf{\tilde{x}}:=\left(
\begin{array}
[c]{c}%
\operatorname{Im}\alpha\\
-\operatorname{Re}\alpha
\end{array}
\right)  ,~\mathbf{\tilde{y}}:=\left(
\begin{array}
[c]{c}%
\operatorname{Im}\beta\\
-\operatorname{Re}\beta
\end{array}
\right)  .
\]

In this notation the displacement becomes $\hat{D}(\mathbf{\tilde{x}}%
)=\exp\left(  i\mathbf{\hat{x}}^{T}\mathbf{\tilde{x}}\right)  $ and the
characteristic function of $\rho_{A|\mathbf{k}}$ reads%
\begin{equation}
\chi_{A|\mathbf{k}}(\mathbf{\tilde{x}}):=\mathrm{Tr}[\rho_{A|\mathbf{k}}%
\hat{D}(\mathbf{\tilde{x}})]=\frac{1}{4\pi p(\mathbf{k})}\chi\lbrack\hat
{O}_{\mathbf{k}}](\mathbf{\tilde{x}}), \label{chiatr}%
\end{equation}
where $\chi\lbrack\hat{O}_{\mathbf{k}}](\mathbf{\tilde{x}})$ is the
characteristic function of $\hat{O}_{\mathbf{k}}$. Using the partial-trace
rule we get%
\begin{equation}
\chi\lbrack\hat{O}_{\mathbf{k}}](\mathbf{\tilde{x}})=\int_{\mathbb{R}^{2}%
}\frac{d^{2}\mathbf{\tilde{y}}}{\pi}\chi(\mathbf{\tilde{x}},-\mathbf{\tilde
{y}})\chi_{0}(\mathbf{\tilde{y}}), \label{ptr}%
\end{equation}
where $\chi_{0}(\mathbf{\tilde{y}}):=\mathrm{Tr}[\rho_{0}(\mathbf{k})\hat
{D}(\mathbf{\tilde{y}})]=e^{-\frac{1}{2}\mathbf{\tilde{y}}^{T}\mathbf{V}%
_{0}\mathbf{\tilde{y}}-i\mathbf{k}^{T}\mathbf{\tilde{y}}}$ and
\begin{align*}
\chi(\mathbf{\tilde{x}},\mathbf{\tilde{y}})  &  :=\mathrm{Tr}[\rho_{AB}\hat
{D}(\mathbf{\tilde{x}})\otimes\hat{D}(\mathbf{\tilde{y}})]\\
&  =e^{-\frac{1}{2}\left(  \mathbf{\tilde{x}}^{T}\mathbf{A\tilde{x}+\tilde{y}%
}^{T}\mathbf{B\tilde{y}+2\tilde{y}}^{T}\mathbf{C}^{T}\mathbf{\tilde{x}%
}\right)  -i\mathbf{\bar{x}}_{A}^{T}\mathbf{\tilde{x}-}i\mathbf{\bar{x}}%
_{B}^{T}\mathbf{\tilde{y}}}.
\end{align*}

By solving the Gaussian integral~\cite{GaussINT} in Eq.~(\ref{ptr}) and
replacing in Eq.~(\ref{chiatr}), we obtain
\[
\chi_{A|\mathbf{k}}(\mathbf{\tilde{x}})=\exp\left(  -\frac{1}{2}%
\mathbf{\tilde{x}}^{T}\mathbf{V}_{A|\mathbf{k}}\mathbf{\tilde{x}%
-}i\mathbf{\bar{x}}_{A|\mathbf{k}}^{T}\mathbf{\tilde{x}}\right)  ,
\]
where the statistical moments, $\mathbf{\bar{x}}_{A|\mathbf{k}}$ and
$\mathbf{V}_{A|\mathbf{k}}$, are those in Eqs.~(\ref{xcond_for})
and~(\ref{CMcond_for}).


\begin{thebibliography}{99}
\bibitem {Sum}Note that the sum can be over a continuous parameter, in which
case it must be interpreted as an integral.
\bibitem {NOTEkw}By applying the Koashi-Winter relation~\cite{KWrel} to the
pure state $\Phi_{ABE}$, we may write $S_{\min}(A|B)=\mathcal{F}(A,E)$, where
$\mathcal{F}(A,E)$ is the entanglement of formation of the two-mode Gaussian
state $\rho_{AE}=\mathrm{Tr}_{B}(\Phi_{ABE})$. If $\rho_{AE}=\rho_{EA}$, then
$\mathcal{F}(A,E)$ can easily be computed as Gaussian entanglement of
formation~\cite{GEOF} and this calculation confirms that $S_{\min}(A|B)$ can
be achieved by Gaussian POVMs.
\bibitem {Units}Here we use $\hbar=2$, so that the shot-noise variance of the
vacuum is equal to one.
\bibitem {Invariance}Any single-mode Gaussian channel can be dilated into a
Gaussian unitary involving the input mode $i$ plus two environmental modes $1$
and $2$. Globally the transformation of the quadratures $\mathbf{\hat{x}}%
^{T}=(\hat{q}_{i},\hat{p}_{i},\hat{q}_{1},\hat{p}_{1},\hat{q}_{2},\hat{p}%
_{2})$ is an affine map $\mathbf{\hat{x}}\rightarrow\mathbf{\hat{x}}^{\prime
}=\mathbf{S\hat{x}}+\mathbf{D}$, where $\mathbf{S}$ is a symplectic matrix and
$\mathbf{D}$ is a displacement vector associated with the channel. Now any
displacement of the input mode $\mathbf{\hat{x}}\rightarrow\mathbf{\hat{x}+d}%
$, with $\mathbf{d}^{T}=(d_{q},d_{p},0,0,0,0)$ will be mapped into a
corresponding (different) displacement of the output mode $\mathbf{\hat{x}%
}^{\prime}\rightarrow\mathbf{\hat{x}}^{\prime}\mathbf{+Sd}$. Since entropy is
invariant under unitaries, it is not affected by these output displacements
and, therefore, by the input ones.
\end{thebibliography}
\begin{thebibliography}{9999}
\bibitem {Proof}In order to prove Eq.~(\ref{chiOB}), we first express the
states in terms of their characteristic functions
\begin{align*}
\rho_{AB}  &  =\int_{\mathbb{C}^{2}}\frac{d^{2}\alpha d^{2}\beta}{\pi^{2}}%
\chi(\alpha,\beta)\hat{D}(-\alpha)\otimes\hat{D}(-\beta).\\
\rho_{0}  &  =\int_{\mathbb{C}}\frac{d^{2}\beta}{\pi}\chi_{0}(\beta)\hat
{D}(-\beta),
\end{align*}
Then, we may write%
\begin{align*}
\hat{O}  &  =\int_{\mathbb{C}^{3}}\frac{d^{2}\alpha d^{2}\beta^{\prime}%
d^{2}\beta}{\pi^{3}}\mathrm{Tr}\left[  \hat{D}(-\beta^{\prime})\hat{D}%
(-\beta)\right] \\
&  \times\chi(\alpha,\beta^{\prime})\chi_{0}(\beta)\hat{D}(-\alpha).
\end{align*}
Since$\ \mathrm{Tr}[\hat{D}(-\beta^{\prime})\hat{D}(-\beta)]=\pi\delta
^{(2)}(-\beta^{\prime}-\beta)$, we get%
\[
\hat{O}=\int_{\mathbb{C}}\frac{d^{2}\alpha}{\pi}\left[  \int_{\mathbb{C}}%
\frac{d^{2}\beta}{\pi}\chi(\alpha,-\beta)\chi_{0}(\beta)\right]  \hat
{D}(-\alpha),
\]
where the term inside the square brackets is $\chi\lbrack\hat{O}](\alpha)$.
\bibitem {GaussINT}For any $n\times n$ real symmetric and positive-definite
matrix $\mathbf{Q}$, we may write%
\[
\int_{\mathbb{R}^{n}}d^{n}\mathbf{x}~e^{-\mathbf{x}^{T}\mathbf{Qx}%
+i\mathbf{d}^{T}\mathbf{x}}=\sqrt{\frac{\pi^{n}}{\det\mathbf{Q}}}%
~e^{-\frac{\mathbf{d}^{T}\mathbf{Q}^{-1}\mathbf{d}}{4}}.
\]
\end{thebibliography}
\end{document}